\documentclass 
{aa}

\usepackage{graphicx}

\begin{document}

\title{Evidence for a 2:3 resonance in Sco X-1 kHz QPOs}

\author{Marek A. Abramowicz\inst{1,2,3} 
\and Tomasz Bulik\inst{4}
\and  Michal Bursa\inst{3,5}
\and  W{\l}odek Klu{\'z}niak\inst{1,4,6}}

\institute{Visiting Professor, 
Institut d'Astrophysique de Paris, 98 bis Blvd. Arago, 75014 Paris, France
\and  Universit\'e Pierre \& Marie Curie (Paris VI), France 
\and Department of Astrophysics, Chalmers University, 412-96
G{\"o}teborg, Sweden, marek@fy.chalmers.se
\and Copernicus Astronomical Centre,
ul. Bartycka 18, 00-716 Warszawa,
Poland, bulik@camk.edu.pl
\and Astronomical Institute, Charles University Prague,
V Hole\v{s}ovi\v{c}k\'ach 2
CZ-180 00  Praha 8,
Czech Republic,
bursa@sirrah.troja.mff.cuni.cz
\and Institute of Astronomy, Zielona G{\'o}ra University,
Lubuska 2, 65-265 Zielona G{\'o}ra, Poland, wlodek@camk.edu.pl}

\offprints{W. Klu\'zniak, \email{wlodek@camk.edu.pl}} 

\date{Received, Accepted}

\abstract{
We find evidence that the two high frequency QPOs in Sco X-1
are, more often than not,  approximately
in the 2:3  frequency ratio familiar from 
studies of black hole candidates (e.g., XTE J1550-564,  Remillard
et al. 2002). This implies that the double kHz QPO phenomenon
in neutron stars has its origin in properties of strong-field gravity
and has little to do with the rotation of a stellar surface
or any magnetic field structure anchored in the star.
}

\maketitle

\section{The larger picture}
\vskip 0.2truecm

Recent X-ray observations showed that in some black holes, as in neutron stars,
high frequency quasi-periodic oscillations (QPOs) come in pairs.
The frequencies of these pairs in the first three 
black hole sources where
they have been observed are 300 Hz and 450 Hz in GRO J1655-40,
42 Hz and 70 Hz in GRS 1915+105, and 184 Hz and 276 Hz in XTE J1550-564
(Strohmayer 2001a,b; Remillard et al. 2002). Clearly,
they are in a 2:3 ratio in two of the sources, and in a 3:5 ratio in the third
(Abramowicz and Klu\'zniak 2001; Klu\'zniak and Abramowicz 2001,2002;
Remillard et al. 2002). In general relativity,
characteristic frequencies scale inversely with the mass.
If one scales the 42 Hz QPO frequency of the
$\sim20M_\odot$ black hole candidate GRS 1915+105
to about one solar mass, i.e., the mass of a neutron star,
one obtains $\sim800\,$Hz, a frequency typical of the lower of the two
``kHz'' QPO frequencies detected in neutron stars.
This in itself suggests that the QPO phenomenon in black holes
and in neutron stars may have the same origin.
We think that the frequency ratios may be even more revealing.

In this paper we examine the published frequency data 
of the well-studied source Sco X-1, a prototype of bright
neutron-star candidates,  to see whether
the two ``kHz'' QPO frequencies may be in the ratio of two low integers,
such as 2:3 or 3:5, as they are in black hole systems.
If such a similarity were well  established between neutron stars
and black holes, it would imply that the ``twin'' high frequency QPOs
originate in  a mechanism operating independently of the presence of
any stellar surface with its co-rotating magnetic fields.
If the high frequency QPOs are fundamentally similar
in neutron stars and in black holes,
they are all likely to be a manifestation of strong-field gravity.

\section{The inner disc radius in neutron-star LMXBs}

 Is the accretion process in neutron star members of
low-mass X-ray binaries (LMXBs) radically different from that in their
black hole counterparts, or are the two processes fundamentally similar?
This question predates the discovery of high frequencies in these systems.
One school of thought held that neutron stars in LMXBs have magnetic
fields and spin rates comparable to that of millisecond 
radio pulsars, which would
imply that the accretion disk is terminated far from the stellar surface,
and that the highest orbital frequencies would be at most two or three
hundred Hertz (within 50 Hz or so of the presumed stellar spin rate).
Others suggested that the magnetic field in these systems
may be so low as to play no dynamical role, and that the discs (presumed to be
geometrically thin) may be terminated close to the central compact object
by effects of strong gravity
much as they are in black holes.
 On this (minority) view,
orbital frequencies in the kilohertz range were expected
(Klu\'zniak, Michelson, and Wagoner 1990).
It would appear that the discovery of kHz quasiperiodic oscillations
in neutron star systems favors this latter view.

The kHz QPOs have fairly similar properties in a wide range of systems,
in particular the highest frequencies have similar values in X-ray
bursters and the much brighter sources, such as Sco X-1
 (see van der Klis, 2000, for a review).
This argues against any special mechanism related to accidental
properties of the system, such as the value of the magnetic field
or of the accretion rate (although
many models based on such finely tuned properties have been proposed).
If, then, the oscillation arises in the accretion flow in the disk,
the orbital frequency sets the scale of expected oscillation
frequencies (Bath 1973). In fact,
the relevant frequency is the meridional epicyclic frequency.
It is identical to the Keplerian frequency for  Newtonian $1/r$ gravity,
but for accretion disks around neutron stars in general relativity 
the meridional epicyclic frequency is
typically somewhat smaller than the orbital frequency.
Observations of a kHz frequency then strongly
disfavor a geometry of accretion in which the maximum orbital frequency
is only a few hundred Hertz. This seems to rule out surface
magnetic dipole
fields in excess of about $10^8\,$G (Klu\'zniak 1998), and tends to favour
the view that the accretion disk in these neutron star systems terminates
close to the marginally stable orbit of general relativity
(Lipunov and Postnov 1984; Klu\'zniak and Wagoner 1985;
Sunyaev and Shakura 1986; Klu\'zniak, Michelson, and Wagoner 1990;
Klu\'zniak and Wilson 1991; Biehle and Blandford 1993;
Zhang, Strohmayer, and Swank 1997;
Kaaret, Ford and Chen 1997; Klu\'zniak 1998; Zhang et al. 1998).

\section{Is the stellar spin frequency directly observed in kHz QPOs?}

Patterson (1979) pointed out that if there is direct interaction
between a structure corotating with the star and a Keplerian accretion disk,
a QPO at the beat frequency $|1/P_* -1/P_{\rm K}|$ may result. Here
$P_*$ is the period of stellar rotation and $P_{\rm K}$ the period 
of orbital motion. Patterson was discussing the white dwarf system
AE Aqu, and the co-rotating structure was an ``illuminating searchlight.''

After the discovery of $\sim40\,$Hz QPOs in LMXBs, this idea
was adapted to a neutron star rotating
at a presumed frequency of $\sim160\,$Hz and beating
close to a presumed Alfven radius against a $\sim200\,$Hz 
Keplerian frequency at the inner edge of an accretion disk
(Alpar and Shaham 1985),
and given the name ``beat-frequency model'' (Lamb et al. 1985).
But a maximum $200\,$Hz Keplerian frequency
is incompatible with the more recently discovered kHz oscillations
in the same sources, so the name has been duly changed to
``sonic-point model'' with one QPO still corresponding 
to a (greatly increased) beat frequency
and the other peak to the orbital frequency (Miller, Lamb and Psaltis 1998).
It has been noted (e.g., M\'endez et al. 1998; Jonker et al. 2002a)
that, in fact, the difference between the two
observed high frequencies in neutron star LMXBs is not constant,
as it presumably should be in the beat frequency model.
Accordingly, other models were proposed in which the interaction
between a Keplerian disk and the stellar magnetosphere 
(``co-rotating structure'') gave rise to  frequencies
in greater concordance with the
observations (Osherovich and Titarchuk 1999).

Alas! The discovery of double
high frequency QPOs also in black hole candidates (Strohmayer 2001a,b;
Miller et al. 2001; Remillard et al. 2002) has  cast a deep shadow over the
notion that there may be  a direct relationship between stellar spin and 
the presence of two high frequency QPOs.
After all, black holes do not have a surface 
and their spin frequency cannot be directly observed.

The general belief seems to be that the distribution
of frequency ratios is fairly uniform, as it is for the source GX5-1
(Fig.~1).
Here, we note that this is indeed the expectation
in any model, where one frequency varies
freely over some range, and the second differs from the first by a 
nearly constant value. There would be no reason for the
observed frequency ratio to cluster about any particular value,
much less a value equal to the ratio of two integers---in the ``co-rotating
structure''
model the ratio between the constant stellar rotation rate
and the orbital frequency is  accidental and cannot be the
ratio of two small integers.
Even stranger would be a clustering about
a value close to that observed in black hole systems, e.g., a 2:3 ratio.
And yet, such a clustering is exactly what we find in Sco~X-1
(Fig.~2).

One weakness of our argument is the uncertain nature
of Sco X-1. Not being an X-ray burster,
Sco X-1 enjoys the status of a neutron star candidate.
We have performed a similar analysis of other neutron star sources,
and the peak at about 2/3 ratio
is present in those data as well (Fig.~3).

\section{Evidence for the 2:3 frequency ratio in Sco X-1}

The high frequency QPOs of Sco X-1 have been discussed in detail
by van der Klis et al. 1997. In particular, the frequencies of the two
QPOs have been presented for a number of separate continuous observations
of 1 to 3 ks each. The frequency, $f_2$, of the upper QPO ranged from about
850 Hz to slightly less than 1100 Hz. The other QPO had a frequency,
$f_1$, lower by about 300 Hz to 200 Hz, respectively. As usual, the difference
$f_2-f_1$ was clearly anticorrelated with the frequency $f_2$.

We have computed the ratio $f_1/f_2$ for each observation
of Sco X-1 reported in van der Klis et al. 1997. 
We find it to be in a much
narrower range than that of the combined data of all neutron 
star sources
(Fig.~3).

For the Sco X-1 data considered, $0.64<f_1/f_2<0.8$.
We have prepared a histogram of the distribution of this QPO frequency
ratio, by binning the ratio in suitable intervals (Fig.~2). 
It is at once evident that
the distribution is not uniform, it is instead
sharply peaked about a central value fairly close to a 2:3 ratio
(indicated in the figure by a vertical dashed line).
This conclusion is not affected by the bin width,
and is fairly representative of the complete sample.

We are not aware of any observational bias which would favor a selection of
those data in which the QPO frequency ratio would cluster about 0.67,
neither are the observers (P. Jonker, private communication).

\section{Statistical significance of the peaks in the frequency ratio}
\vskip 0.2truecm

To quantify our claim of clustering about a preferred value
of the frequency ratio of ``kHz'' QPOs in Sco X-1, 
we have fit to the observed distribution
of frequency ratios a
suitably normalized Lorentzian as a trial probability
distribution,
$$ {dp\over dr} = {N(r_o,\lambda)\over (r-r_o)^2 + \lambda^2 },\eqno(1)$$
where $r_0$ and $\lambda$ are free parameters, and $N(r_o,\lambda)$
is the normalization factor. Here, $r\equiv f_1/f_2$.
We use the maximum likelihood method (Bulik, Klu\'zniak, and Zhang 1999)
to find the probability distribution that best describes the data,
i.e., the most likely value of the two parameters.
Once these are determined, we calculate
the probability that the data has been drawn from this
distribution using the Kolmogorov Smirnov test.
The likelihood function is the product of probabilities 
of measuring each individual frequency ratio
$
{\cal L} (r_0,\lambda) = \Pi_i ({dp/dr}|_{r_i}).
$

We searched the two dimensional parameter space for 
the best values of $r_0$ and  $\lambda$ and found a strong peak
at $r_0=0.68$ and $\lambda=0.018$. The Kolmogorov Smirnov
test shows that the probability that the data has been 
drawn from the distribution described by eq. (1) with these two
values of the parameters is $p=0.39$. This confirms that 
our initial choice of the shape of the distribution (Lorentzian)
is close to the original distribution of the ratios,
and that the probability distribution maximizing
the likelihood function is acceptable.

In Figure 2, we superpose  on the histogram
this best-fit Lorentzian distribution (smooth curve),
normalized to the total number of events. 
The offset of the peak from the value 2/3 is statistically significant.
A fit with a forced value of the centroid at $r_0=2/3$ yields
a slightly broader distribution with $\lambda=0.021$,
but the K-S probability is unacceptable, at $p= 7.1\times 10^{-4}$.

The data of fig.~2, here binned in intervals of  width $d\,r=0.02$,
exhibit an additional peak around the
value of $r\approx 0.78 $.
We have also tried fitting two Lorentzians.
The best fit values  are $r=0.679$ and 0.7763 for the centroids,
and $\lambda=0.01369$ and 0.0049 for the widths of the two peaks, respectively.
 Although  inclusion of
a second Lorentzian improves the fit (K-S probability of 0.801),
 from the statistical
point of view the second peak is not strongly required by the data,
as the K-S test probability 
is already acceptable with just one Lorentzian.

\begin{figure}
\centering
\includegraphics[width=8.5cm]{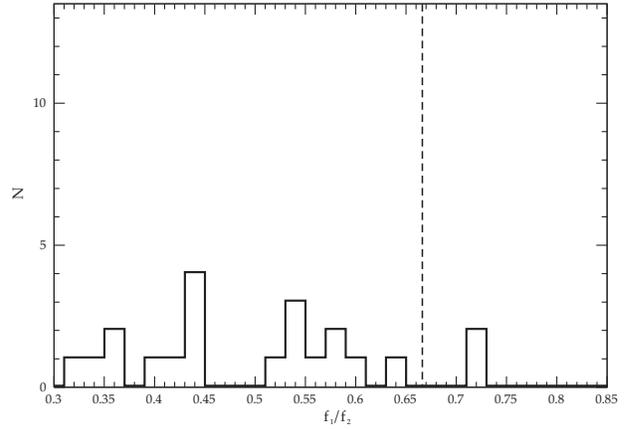}
\caption{The observed distribution of frequency ratio of the two kHz QPOs 
of GX5-1. Also shown is the location of the 2:3 ratio (dashed line).
The data is from Jonker et al. 2002b.}
\label{gx5}
\end{figure}
\begin{figure}
\centering
\includegraphics[width=8.7cm]{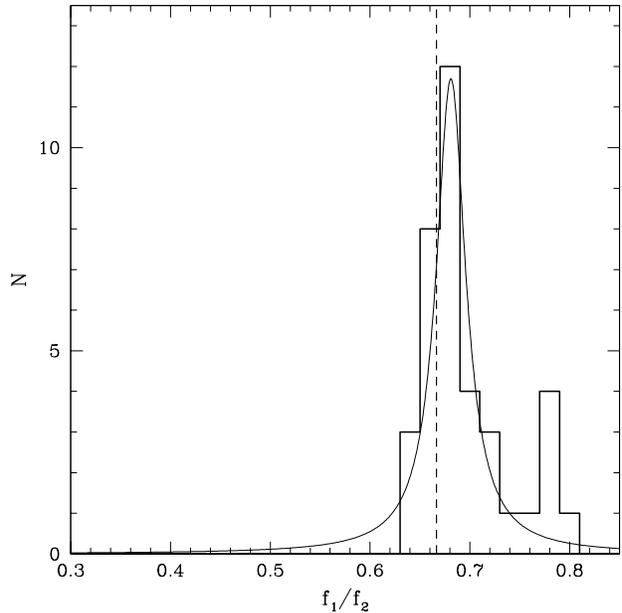} 
\caption{The same for Sco X-1. The data is from van der Klis et al. 1997.
Also shown is the best fitting Lorentzian (eq. [1]),
note the slight offset of the peak from the value 2/3.}
\label{sco}
\end{figure}

A careful analysis of
Fig. 3 in van der Klis et al. 1997, especially panel d),
suggests that the second peak may be real---the power ratio of the two QPOs
departs from the general trend when $f_1/f_2$ becomes greater than 0.75
(i.e., when $f_2> 1050\,$Hz).
From the data (Fig.~2 of this work and Fig.~3 of van der Klis et al. 1997),
it would appear that when the QPO frequency ratio is $f_1/ f_2=2/3$
(to  within a few percent), the upper QPO has about
twice as much power as the lower QPO, and the upper QPO power steadily
decreases with increasing frequency ratio $f_1/f_2$,
until $f_1/ f_2\approx 0.75$, when the upper QPO again becomes twice,
and even three times,
as powerful as the lower QPO.

Recent observations
of the burster 4U 1636-53, support the reality of both peaks
in Fig.~2. The frequency difference between the QPOs in 4U 1636-53 
seems to undergo an abrupt transition from a peak separation
of about 330~Hz to about 250 Hz 
(Jonker et al. 2002a, DiSalvo et al. 2002).
On one side of the transition the frequency ratios
are 0.662, 0.680, 0.685, 0.725, i.e., they
 may well
have been sampled from the 2/3 peak in the distribution of Sco X-1;
on the other side, the ratios vary between 0.75 and 0.78.

\section{Interpretation of the results}

Figs.~1, 2, and 3 make it clear that, unlike GX5-1,
most neutron star sources do not
show a flat distribution of kHz QPO frequency ratios.
It is hard to escape the conclusion
that a resonance is responsible for the distribution of frequency
ratios in Sco X-1. The 2/3 value, identical to the ratio seen
in the black hole candidates GRO J1655-40 and XTE J1550-564
(whose measured masses are $\sim6M_\odot$ and $\sim8M_\odot$)
is a very strong indication that the high frequency QPOs
are a manifestation of strong-field gravity, and not of 
the presence of a stellar surface
with any co-rotating structure.

Our argument about the role played by strong field-gravity
in forming the high frequency QPOs is based on the similarity
between neutron stars and black holes, for this
we need to assume that the microquasars
GRO J1655-40 and XTE J1550-564 are in fact accreting black holes.


\begin{figure}
\centering
\includegraphics[width=8.5cm]{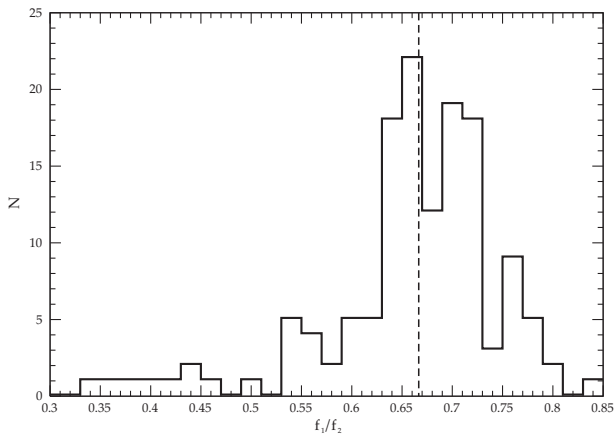} 
\caption{The combined distribution of frequency ratios
for
4U~0614+09,
4U~1608-52,
4U~1636-53,
4U~1702-43,
4U~1705-34,
4U~1728-34,
4U~1735-44,
4U~1820-30,
4U~1915-05,
Cyg~X-2,
GX~17+2,
GX~340+0,
GX~349+2 (Sco X-2),
KS~1731-26,
and~XTE J2123-058.
}
\label{histogram}
\end{figure}

To make the argument secure, one needs a theory of the origin
of high frequency QPOs able to accomodate both
the sharp frequency ratios in the observed black holes
and the broader distribution observed in Sco X-1 and some other
neutron star sources. It must explain a preference for a 2/3 ratio
of the high frequency QPOs and the offset between the peak of the
distribution of Fig.~2 and the value 2/3.
We note that all these facts
seem to be in accord with the theory of epicyclic parametric resonance
in relativistic accretion disks. On that theory
(Klu\'zniak and Abramowicz 2002), the higher frequency
QPO occurs at the frequency of meridional oscillations, $\Omega_z$,
in that orbit in which the radial epicyclic frequency is
 $\Omega_r=(2/3)\Omega_z$. For the Schwarzschild metric this occurs
at $r_{2:3}=5.4r_g=16.2\,{\rm km}/M_\odot$, and comparison with
the observed frequency of 900 Hz in Sco X-1
implies that $M=1M_\odot$, with modestly higher values
for rotating neutron stars.

The origin of the second peak of frequency ratios, at $\sim7/9$,
 remains a puzzle,
as does the flat distribution of frequency ratios seen in
GX5-1 and some other sources.
Detailed comparison of observations and the theory of parametric 
epicyclic resonance for realistic
space-time metrics of rapidly rotating neutron stars will be the subject
of another paper.

\begin{acknowledgements} 
 We thank Peter Jonker, Michiel van der Klis and Jean-Pierre
Lasota for valuable discussion. 
 This work has been funded in part through the KBN grants
 5P03D01721 and 2P03D02117, by CNRS, and by the Swedish Research Council.
 MAA and WK thank Institut d'Astrophysique de Paris
 for generous support and hospitality.
\end{acknowledgements}

\end{document}